\documentclass[10pt,letterpaper]{article}
\usepackage{opex3}
\usepackage{graphicx}
\usepackage{subfigure}
\usepackage{float}
\usepackage{bm}
\usepackage{mathcomp}

\def\be{\begin{equation}}
\def\ee{\end{equation}}
\def\bea{\begin{eqnarray}}
\def\eea{\end{eqnarray}}
\def\ba{\begin{array}}
\def\ea{\end{array}}

\def\0{$\Gamma_0$}

\def\part{\partial}

\begin{document}

\title{Superlensing properties of one-dimensional dielectric photonic crystals}

\author{Salvatore Savo$^{1}$, Emiliano Di Gennaro$^{2}$ and Antonello Andreone$^{1}$}

\address{$^{1}$CNR - INFM ``Coherentia'' and Physics Department, University ``Federico II'', Napoli, Italy\\
$^{2}$CNISM and Physics Department, University ``Federico II'', Napoli, Italy}

\email{savo@na.infn.it} 



\begin{abstract*}
We present the experimental observation of the superlensing effect in a slab of a one-dimensional photonic crystal made of tilted dielectric elements. We show that this flat lens can achieve subwavelength resolution in different frequency bands. We also demonstrate that the introduction of a proper corrugation on the lens surface can dramatically improve both the transmission and the resolution of the imaged signal.
\end{abstract*}



\section{Introduction}

The very interesting optical properties shown by photonic crystals (PCs) make them extremely attractive and suitable for a large variety of applications, like waveguides, beam splitters, wavelength multiplexers, resonators, collimators and many others. Since the publication of the perfect superlens having sub-wavelength resolution \cite{pen}, many efforts have been spent both theoretically \cite{noto,luo1,luo2} and experimentally \cite{par} to study the negative index properties of PCs. Most of the research done till now has been focused on two-dimensional (2D) PCs, with a few theoretical works referring to the features of one-dimensional (1D) PCs realized by alternating dielectrics with either metals \cite{fan1,fan2} or negative index materials    \cite{shi}. Only recently, further investigation has been conducted to study the imaging properties of flat lenses realized with 1D PCs made of dielectric materials only \cite{he}. It has been proved both theoretically and numerically that it is possible to achieve off-axis sub-wavelength imaging resolution (FWHM = 0.16$\lambda$ - 0.20$\lambda$) with a slab made of tilted alternating layers of different permittivity. The advantage of a lens composed of dielectric elements is in the drastic reduction of losses that normally come from the metal. This renders such structures particularly suitable for applications in the telecommunication wavelength regime.

In this paper, we first show by means of numerical simulations the improvements produced in the lens performances introducing a simple corrugation on both slab surfaces, in terms of both the imaging quality and the transmission efficiency. We also investigate how the frequency windows where superlensing is achieved depend on the relative widths and on the index contrast of the dielectric layers.

Then, we carry out an experimental investigation in the microwave regime showing for the first time, to the best of our knowledge, the sub-wavelength focusing properties of this new type of 1D flat lens with corrugation. We perform point imaging measurements using two different lenses, having the same lattice properties but different index contrast. The first lens consists of alternating layers of alumina ($Al_2$$O_3$, $\epsilon_r$ = 8.6) and air whereas the second one is made of alumina and plexiglas ($\epsilon_r$ = 2.5).

\section{Band structure analysis}

The focusing properties of a flat lens made of tilted dielectric elements can be easily understood from the analysis of its Equi-Frequency Contours (EFCs), calculated using the transfer matrix method as in \cite{feng}.

In the following we will assume TE polarization only, that is the electric field E perpendicular to the plane of incidence.

\begin{figure}[htbp!]

\centering

\includegraphics[width=8.5cm]{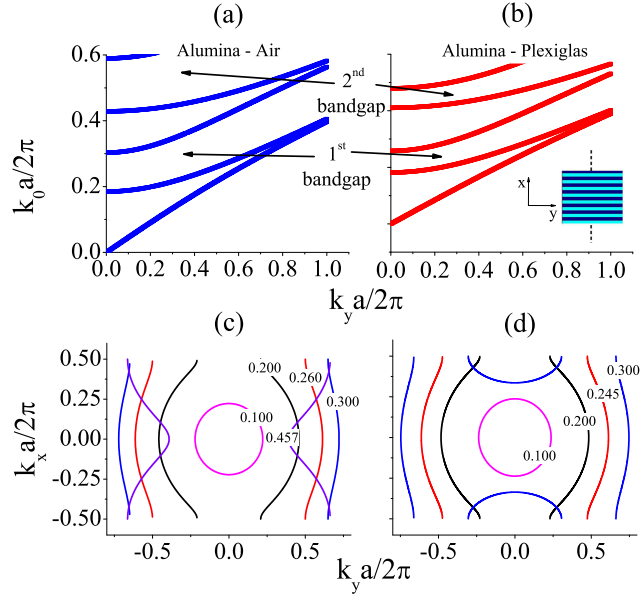}

\caption[]{(Color online) Band structure along the off-axis direction for a 1D PC with a unit cell realized using alumina and air (a) or alumina and plexiglas (b) respectively. (c) and (d): EFCs at different frequencies for the same structures.}

\label{BS}

\end{figure}

Consider a 1D photonic crystal of infinite extension (see inset in Fig. \ref{BS}(b)) consisting of two different alternating dielectric materials having width $\emph{W}_1$ and $\emph{W}_2$ and permittivity $\epsilon_1$ and $\epsilon_2$ respectively. Assume that the layers are parallel to the \emph{y}-axis with a period $\emph{a} = \emph{W}_1 + \emph{W}_2$ along the \emph{x} direction.

Figures \ref{BS}(a)-(d) show the band structures and EFCs for the two different cases that we have experimentally studied in this work: the alumina-air (Figs. \ref{BS}(a) and \ref{BS}(c)) and the alumina-plexiglas case (Figs. \ref{BS}(b) and \ref{BS}(d)) respectively.

\begin{figure}[htbp!]

\center{\includegraphics [angle=0, width=14cm]{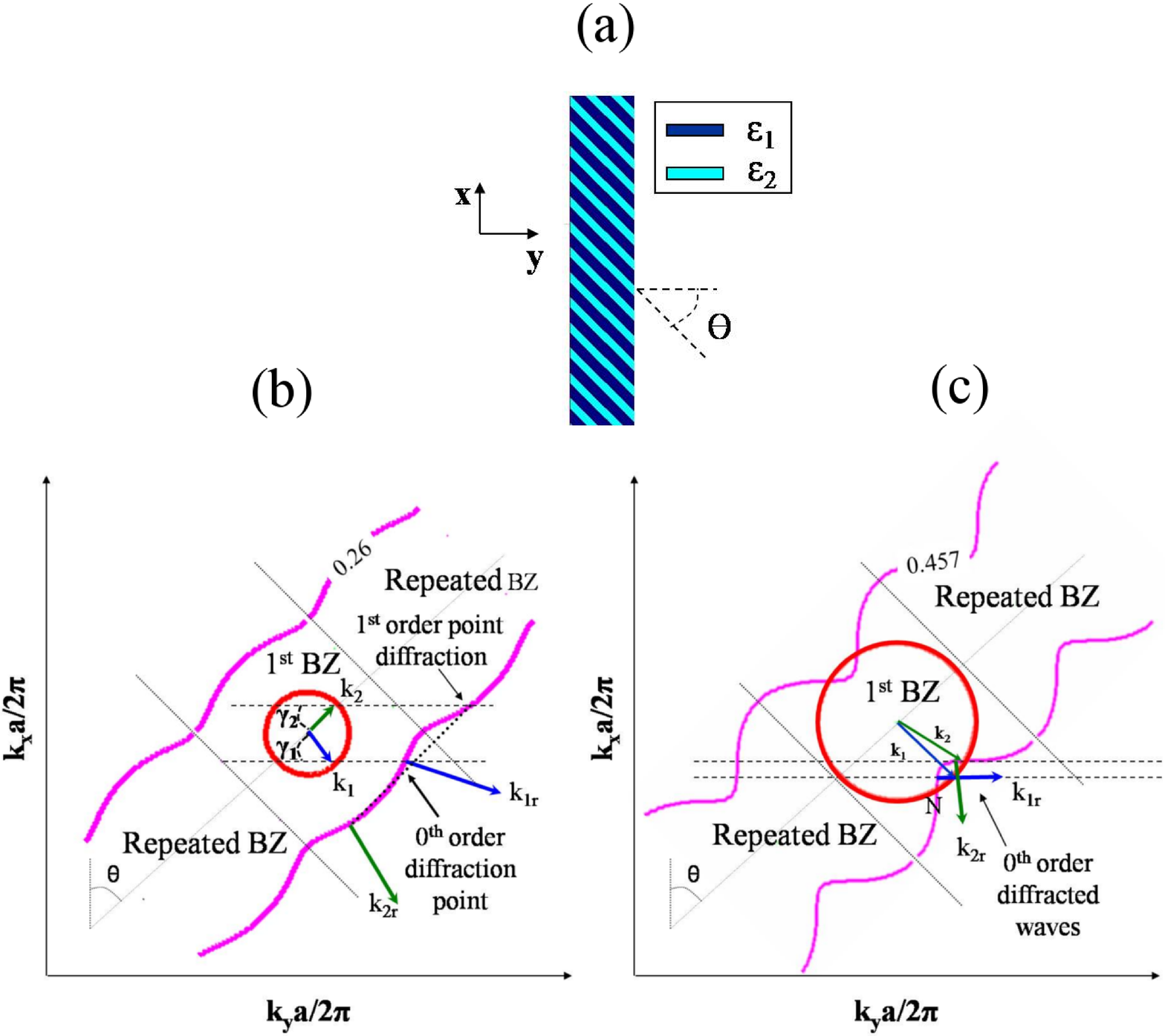}}

\caption{(Color online) (a) Sketch of the slab obtained rotating, by an angle $\theta$, a one dimensional PC having a unit cell  made of two different dielectric materials $\epsilon_1$ = 8.6 and $\epsilon_2$ = 1. (b) and (c) equi-frequency contours at $\omega=0.260$ ($1^{st}$ band) and $\omega=0.457$ ($2^{nd}$ band) respectively, plotted in the first and in the repeated Brillouin zone, relatively to the crystal in (a); $k_1$ and $k_2$ are two wave vectors impinging the air-PC interface at two generic different angles $\gamma_1$ and $\gamma_2$, whereas $k_{1r}$ and $k_{2r}$ are the wave vectors diffracted inside the crystal.}

\label{WV}

\end{figure}

When the dielectric layers are all tilted by the same angle $\theta$ with respect to the normal to the lens (see Fig. \ref{WV}(a)), the diffraction inside the structure can give rise to interesting focusing phenomena. The rotation of the layers in the spatial domain implies the same amount of rotation for the EFCs in the \emph{k}-space. In Figs. \ref{WV}(b) and (c) the wave vector diagrams at two different frequencies for a 1D PC, consisting of alumina-air alternating layers  with $\emph{W}_1 = \emph{W}_2 = 0.5\emph{cm}$, are shown. The magenta curves represent the equi-frequency contour for such structure, plotted in the first and repeated Brillouin zone (BZ), whereas the red circle is the air EFC. Curves are plotted at $\omega = 0.260$ (Fig. \ref{WV}(b)) and $\omega = 0.457$ (Fig. \ref{WV}(c)), where $\omega = a/\lambda$  is the normalized frequency. The two arrows inside the red circles in Figs. \ref{WV}(b) and (c) are two generic wave vectors impinging the air-PC interface at different incoming angles $\gamma_1$ and $\gamma_2$. The directions of the diffracted wave vectors inside the PC are determined by imposing the conservation of the tangential component $k_i^{//}=\omega /c \sin{\gamma_i}$, here represented by the construction line drawn as a dashed line. The two wave vectors normal to the PC EFCs, $k_{1r}$ and $k_{2r}$, are the diffracted waves that propagate inside the crystal. They are both directed perpendicularly to the EFC and point away from the source.

There is however a fundamental difference between the diffracted wavevectors depicted in Fig. \ref{WV}(b): $k_{1r}$ corresponds to an intersection between the construction line and the EFC lying in the $1^{st}$ BZ whereas $k_{2r}$ derives from the intersection in the repeated BZ and therefore is subjected to the folding back process. As it is easily seen in the plot of Fig. \ref{WV}(b), $k_{1r}$ and $k_{2r}$ are positively and negatively refracted respectively. This phenomenon is a direct consequence of the equi-frequency contours shape shown by the 1D structure at the chosen normalized frequency.

At $\omega = 0.457$, although the shape of the EFC (see Fig. \ref{WV}(c)) is very different from the one at $\omega = 0.260$, it is still possible to obtain the conditions for negative refraction due to the convex nature of the EFC with only $0^{th}$ order diffracted waves involved.

These remarkable diffraction properties can be exploited to realize superlensing, since using a simple ray-diagram it is easy to show that an appropriately designed slab made of this kind of 1D PC will have an off-axis focusing. The performance of a flat superlens made of tilted dielectric elements has been already presented and discussed numerically by Wang et al. \cite{he}.

The focusing properties strongly depend on the orientation of the surface termination via the rotation angle $\theta$. Off-axis subwavelength focusing is achieved in  \cite{he} for a high index contrast $(Si-SiO_2)$ layered structure, with the best FWHM of 0.164$\lambda$  for  $\theta$ = $44\tcdegree$. In this work we chose $\theta$ = $45\tcdegree$ since this gives us the best resolution for the index contrast structures presented here, consisting of layered elements of alumina-air and alumina-plexiglas.

\section{Focusing properties and effect of surface corrugation}

We also show, by means of Finite-Difference Time-Domain (FDTD) simulations  that the surface corrugation coming from the simple rigid rotation of the dielectric layers can dramatically improve the performance of this flat lens.

The discretization grid for the simulation domain is a/30 along the x and y direction. A point source is located on the slab axis and centered $1cm$ far from the PC interface without corrugation. The source holds the same position also for the case with corrugation.

\begin{figure}[htbp!]

\centering

\subfigure{

\hspace{0.2cm}

\includegraphics[width=5cm]{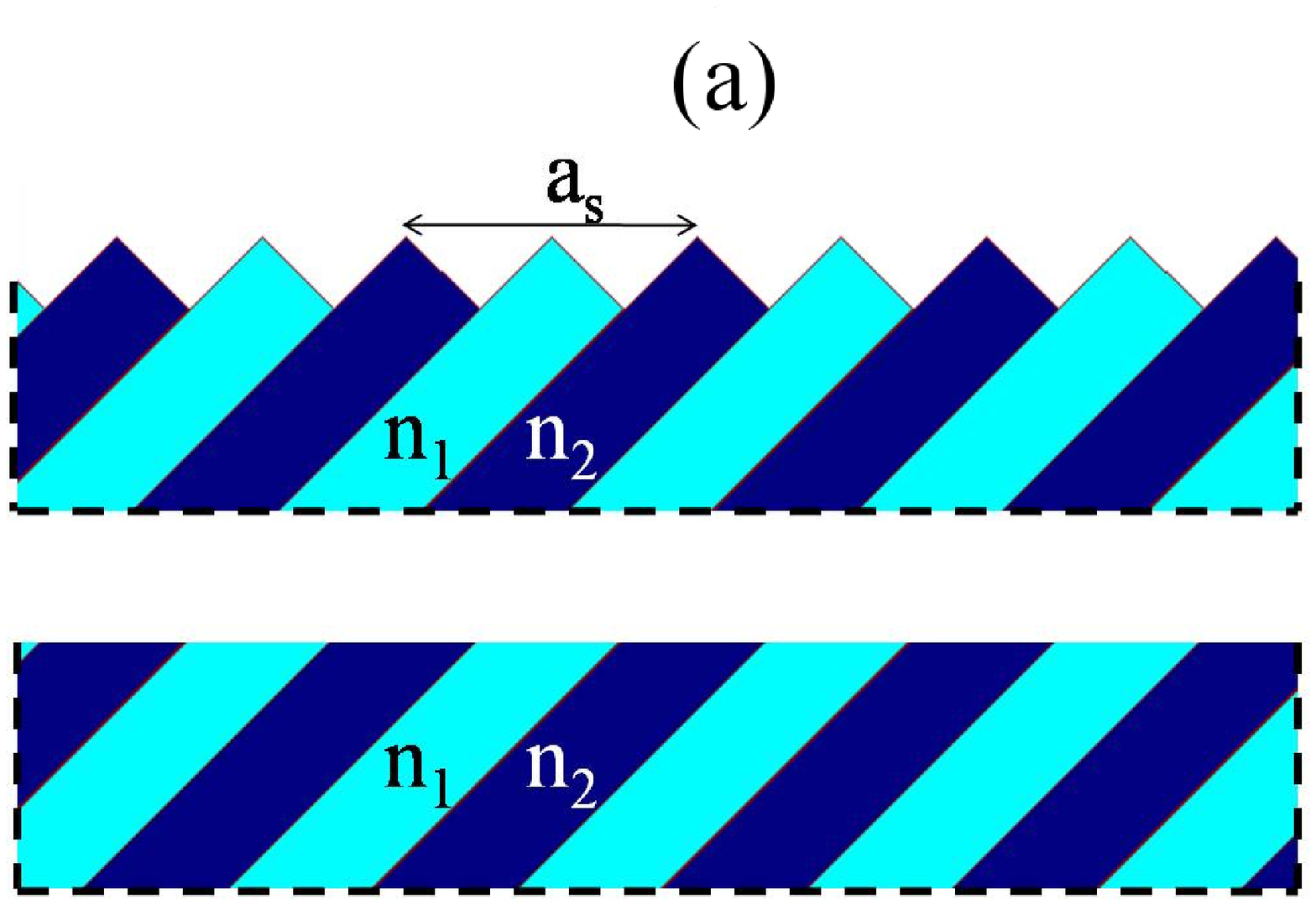}

\hspace{-0.45cm}

\includegraphics[width=4.5cm]{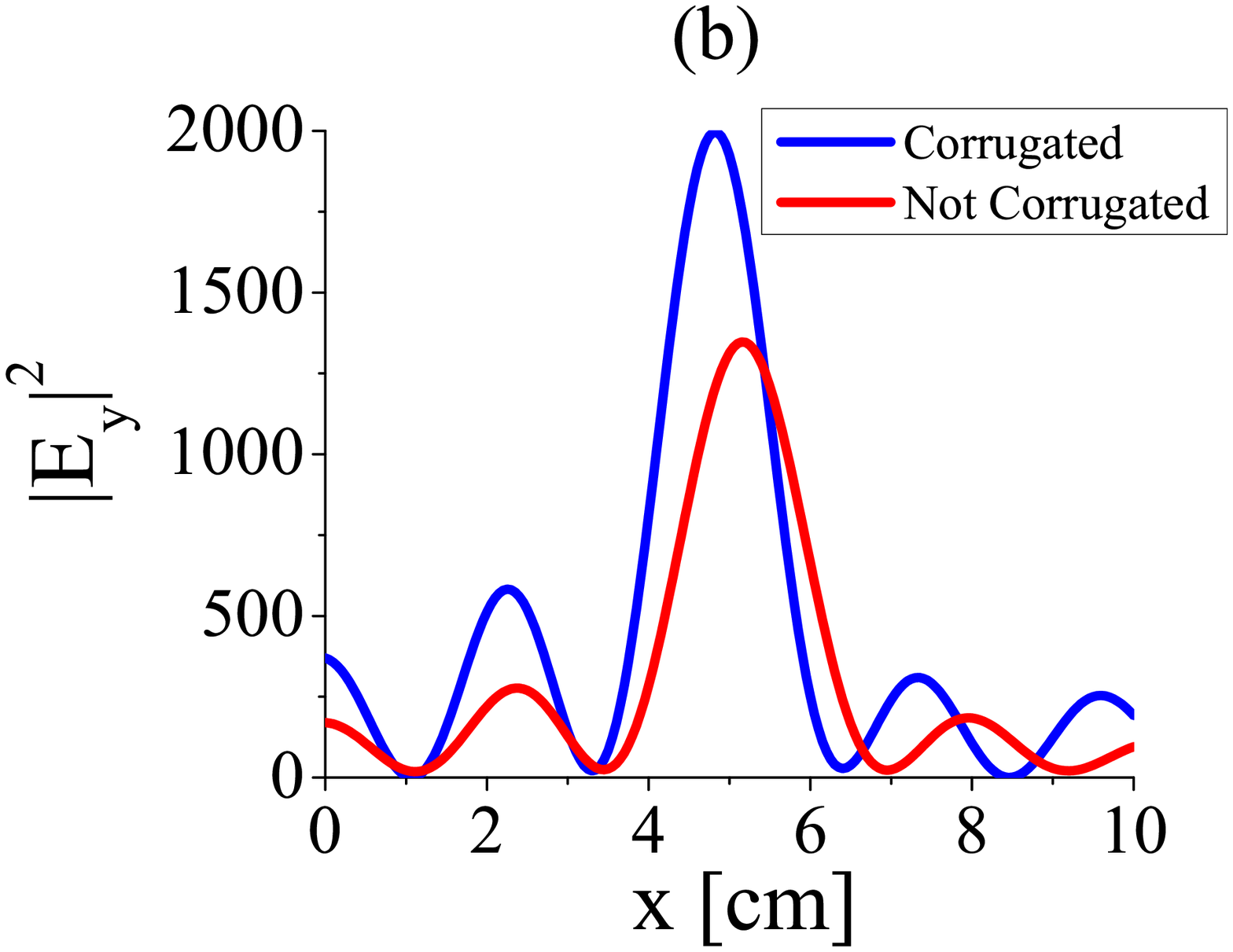}

}

\subfigure{

\includegraphics[width=4.5cm]{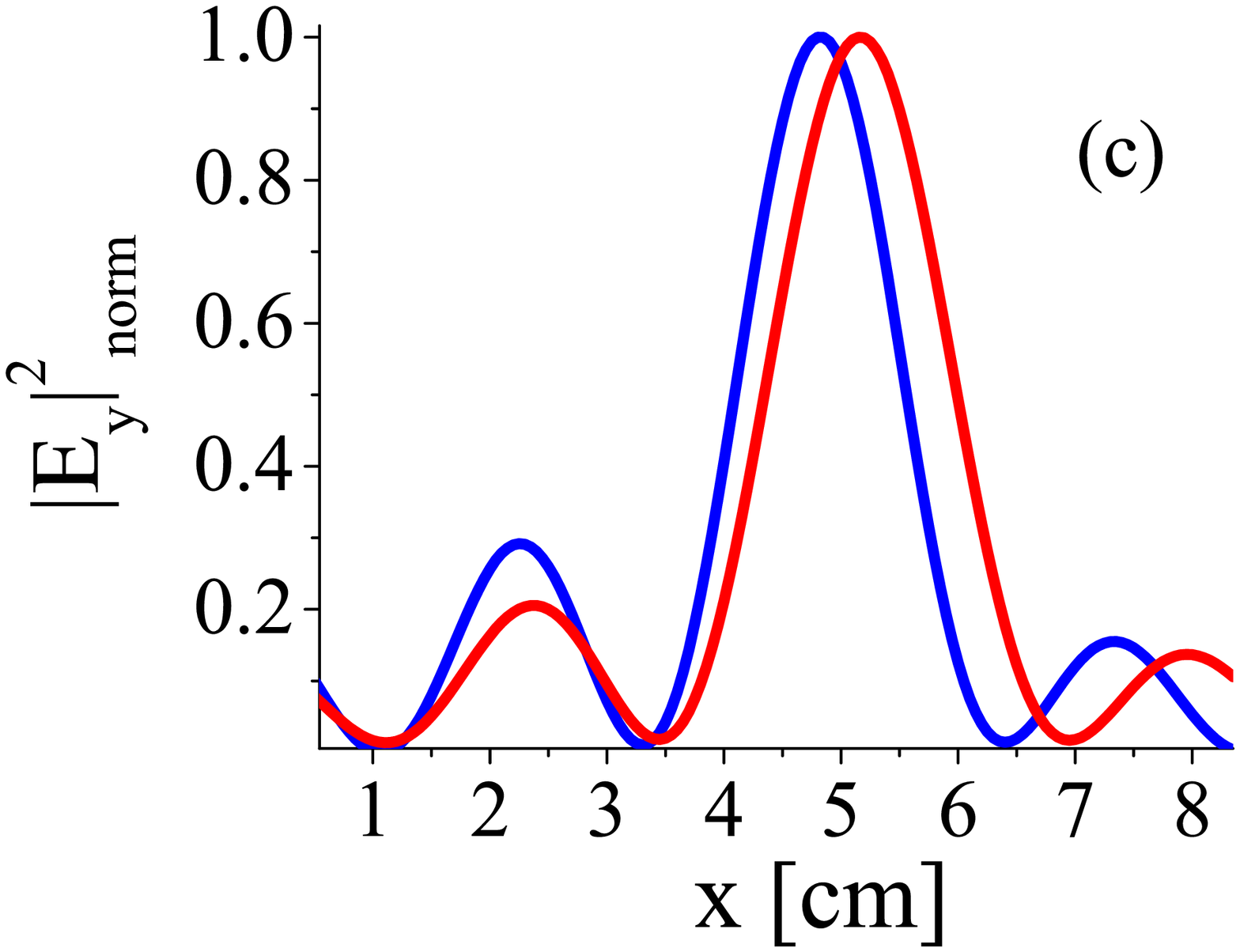}

\hspace{0.7cm}

\includegraphics[width=4.5cm]{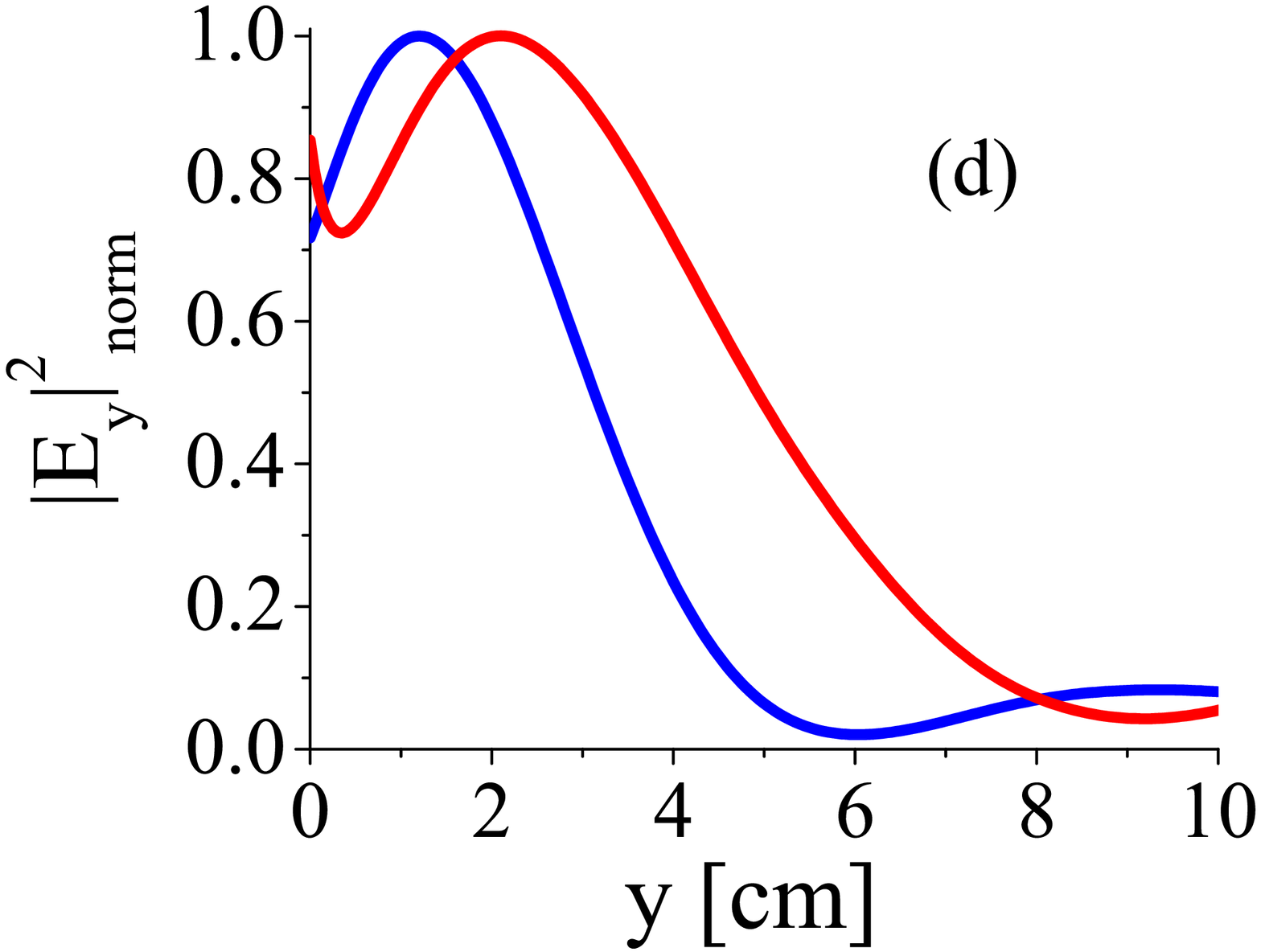}

}

\caption{(Color online) (a) Detail of the photonic crystal surface termination for the case with and without corrugation. (b) Comparison of the focus profiles (calculated via FDTD) produced by the slab with corrugation (blue line) and without corrugation (red line). (c) and (d) Transversal and lateral profiles respectively of the transmitted energy in a normalized scale. All the above profiles are taken along the lines (parallel and perpendicular to the PC surface) where the focused image exhibits its maximum.}

\label{SIM1}

\end{figure}

In Fig. \ref{SIM1}(a) a detail of the lens we simulated ($\epsilon_1$ = 8.6,  $\epsilon_2$ = 2.5), with and without corrugation on both surfaces, is shown. For the case under study, this particular kind of corrugation yields two evident benefits. First, it increases the transmission efficiency (Fig. \ref{SIM1}(b)) by about 50\%, because of the reduced impedance mismatch with the surrounding medium. Then, it produces also a significant improvement in both the transversal and lateral resolution (about 11\% and 36\% respectively), as shown in Figs. \ref{SIM1}(c) and (d).

These enhancements are usually related to the presence of surface states introduced by the surface corrugation \cite{dec}. The benefits to the imaging quality introduced by the surface states are a well known property of photonic crystals \cite{luo2,mea}. It has been proved both numerically \cite{xiao} and experimentally \cite{wang,casse} that choosing the appropriate corrugation can increase the imaging resolution performance of a PC superlens. Furthermore, surface states can be exploited as well to realize beam shaping with using waveguide made of 2D photonic crystal with corrugated surface \cite{mor}.

\begin{figure}[htbp!]

\centering









\includegraphics[width=10cm]{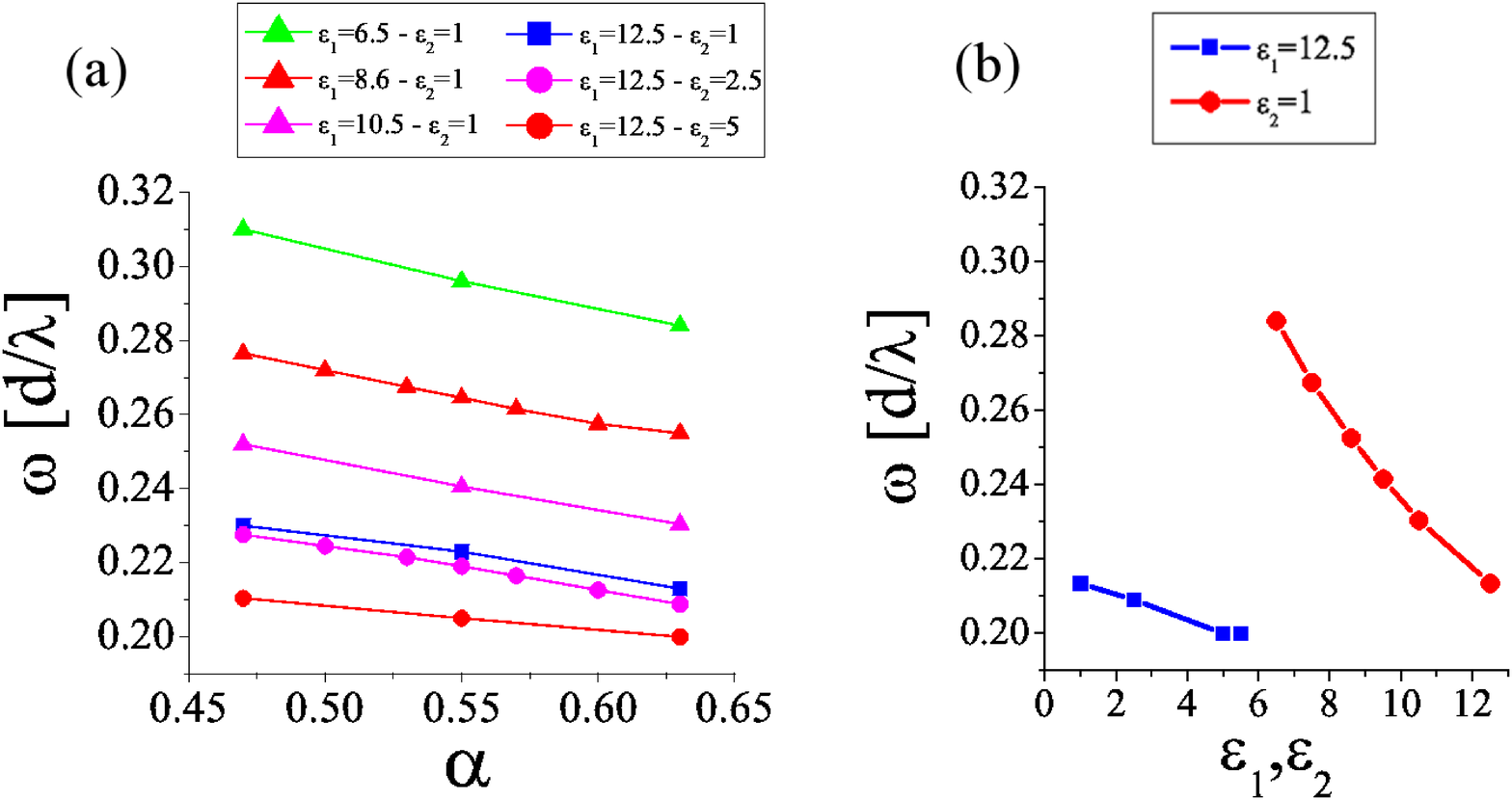}

\caption[]{(Color online) Plot of the normalized frequencies at which slabs made of different materials show the maximum transmission for the focused image as a function of the finite width ratio $\alpha$ (a) and dielectric constants $\epsilon_1$, $\epsilon_2$  (b) of the tilted elements ( $\theta = 45\tcdegree$).}

\label{SIM2}

\end{figure}

We also numerically evaluated how the frequency at which the lenses show the maximum peak intensity for the focused image varies with the finite width ratio $\alpha= \emph{W}_1/ (\emph{W}_1+\emph{W}_2)$. A parametric study for different values of the dielectric constants $\epsilon_1$ and $\epsilon_2$ was carried out. Results show (see Fig. \ref{SIM2}(a)) that within a relatively wide range centered at $\alpha$ = 0.55 the frequency linearly decreases with increasing $\alpha$. Moreover, the curves that fit different sets of data, including the cases of alumina-air and alumina-plexiglas based PCs, are approximately parallel.

The lens performance severely degrades, with no focusing observed at all outside the range [0.47, 0.63]. For a fixed value of $\alpha$, the focusing frequency linearly decreases also by increasing the permittivity of one of the two materials (Fig. \ref{SIM2}(b)). The square-dotted line represents the case where the elements with the highest permittivity are kept constant ($\epsilon_1$=12.5), letting $\epsilon_2$ to vary from 1 to 5.5, whereas the circle-dotted line represents the case  with $\epsilon_2$ = 1, and $\epsilon_1$ is let to vary from 5 to 12.5. We think that it is possible to use these findings as simple ``rule of thumb'' to design a flat 1D PC superlens consisting of tilted dielectric elements.

\section{Measurement setup}

Experimental results are carried out by sandwiching the PCs in an aluminium parallel-plates waveguide terminated with microwave absorbers and measuring the amplitude and phase of the electric field transmitted by the crystals in the image plane. Since the loss tangent of both alumina and plexiglas are extremely small at the frequency of interest for this work ($\tan{\delta_{Al}} < 10^{-3}$, $\tan{\delta_{Pl}} < 10^{-2}$), dielectric losses can be neglected.

A dipole antenna (radius = $0.6mm$) is used as a source, oriented to produce an electric field \emph{z}-oriented (TE mode) and operating in the range of frequencies that spans from $5 GHz$ to $15 GHz$, in order to reproduce the same normalized frequency $a/\lambda$  of the theoretical model. Due to the waveguide characteristics, the TEM mode only can propagate up to $15 GHz$. The maps of the amplitude and phase of the electric field are collected by using a HP8720C Vector Network Analyzer. Another dipole antenna with the same characteristics of the source is used as a detector that moves along the waveguide plane using an x-y step motor. Details of this technique has already been presented in \cite{savo}.

We have conducted point imaging experiments on two different structures using, in the first case a unit cell made of alumina and plexiglas, in the second case alumina and air. It is straightforward that the latter structure comes as the natural consequence of the first one, since it is obtained by the simple removal of all plexiglas layers. For this reason the two PCs have the same lattice properties but a different index contrast. The PC slabs before rotation, are built with a lattice constant $a = 1cm$, length $L = 30cm$ and width $w = 5.65cm$. We chose dielectric elements having the same width, therefore the corresponding layer widths in the unit cell are 0.5\emph{a} for both structures under study.

When all layers are rotated by the same angle $\theta$, the PC can be seen as a 2D structure having lattice constants along \emph{x} and \emph{y} directions given by  $a_x = \emph{a}/\sin{\theta}$  and $a_y = \emph{a}/\cos{\theta}$ respectively, length $L_s = 30 a/\cos{\theta}$  and width  $w_s = w \sin{\theta}+0.5\emph{a}\cos{\theta}$, where the width is defined including the corrugation.

\begin{figure}[h]

\center{\includegraphics [angle=0, width=5cm]{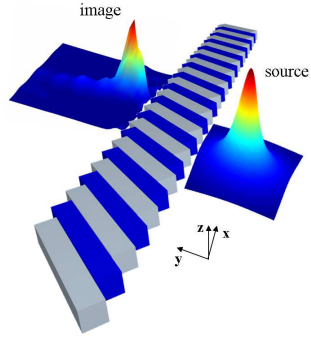}}

\caption{ (Color online) 3D spatial mapping of the point source signal transmitted by the alumina-air slab at $13.7175GHz$.}

\label{PC}

\end{figure}

For the sake of clarity, in Fig. \ref{PC} is shown the 3D map of the measured point source imaged by the slab. More details of the measurements will be given in next paragraph.

\section{Experimental results and discussion}

Measurements are performed scanning an area $10cm$ wide and $10cm$ long adjacent to the PC-air interface, in steps of $2mm$ in both x and y direction. The point source is positioned $1cm$ far from the lens in $x = 0cm$, according to our reference system.

\begin{figure}[htbp!]

\center{\includegraphics [angle=0, width=10cm]{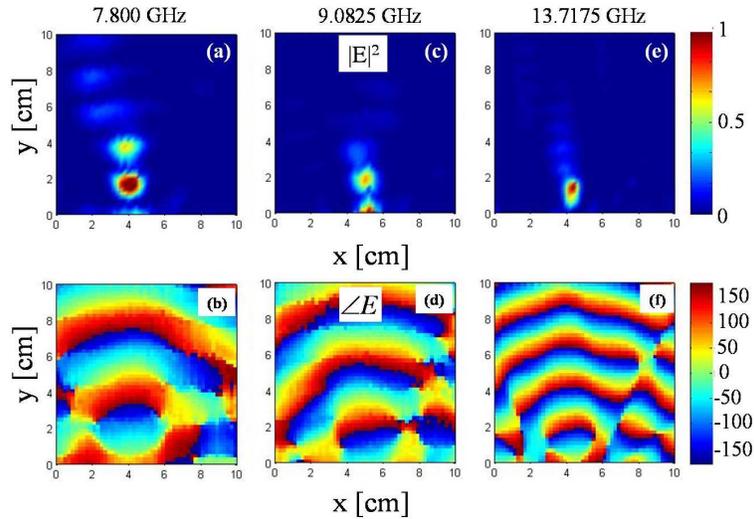}}

\caption{ (Color online) In-plane spatial mapping of the measured electric field intensity ((a), (c), (e)) and phase ((b), (d), (f)) for the alumina-air PC at $7.800GHz$, $9.0825GHz$ and $13.7175GHz$ respectively.}

\label{MEAS2}

\end{figure}

In Figs. \ref{MEAS2}(a)-(f) the complete experimental results for the alumina-air lens are illustrated. This structure exhibits subwavelength imaging in three different bands, $BW_1=[7.6200-7.8200]GHz$, $BW_2=[8.4700-9.2200]GHz$, $BW_3=[13.4500-14.2300]GHz$. The signals imaging are clearly off-axis as evident from the electric field intensity maps in Figs. \ref{MEAS2}(a), (c), (e), and as also predicted by \cite{he}.


\begin{figure}[htbp!]
\centering
\subfigure{
\includegraphics[width=4.5cm]{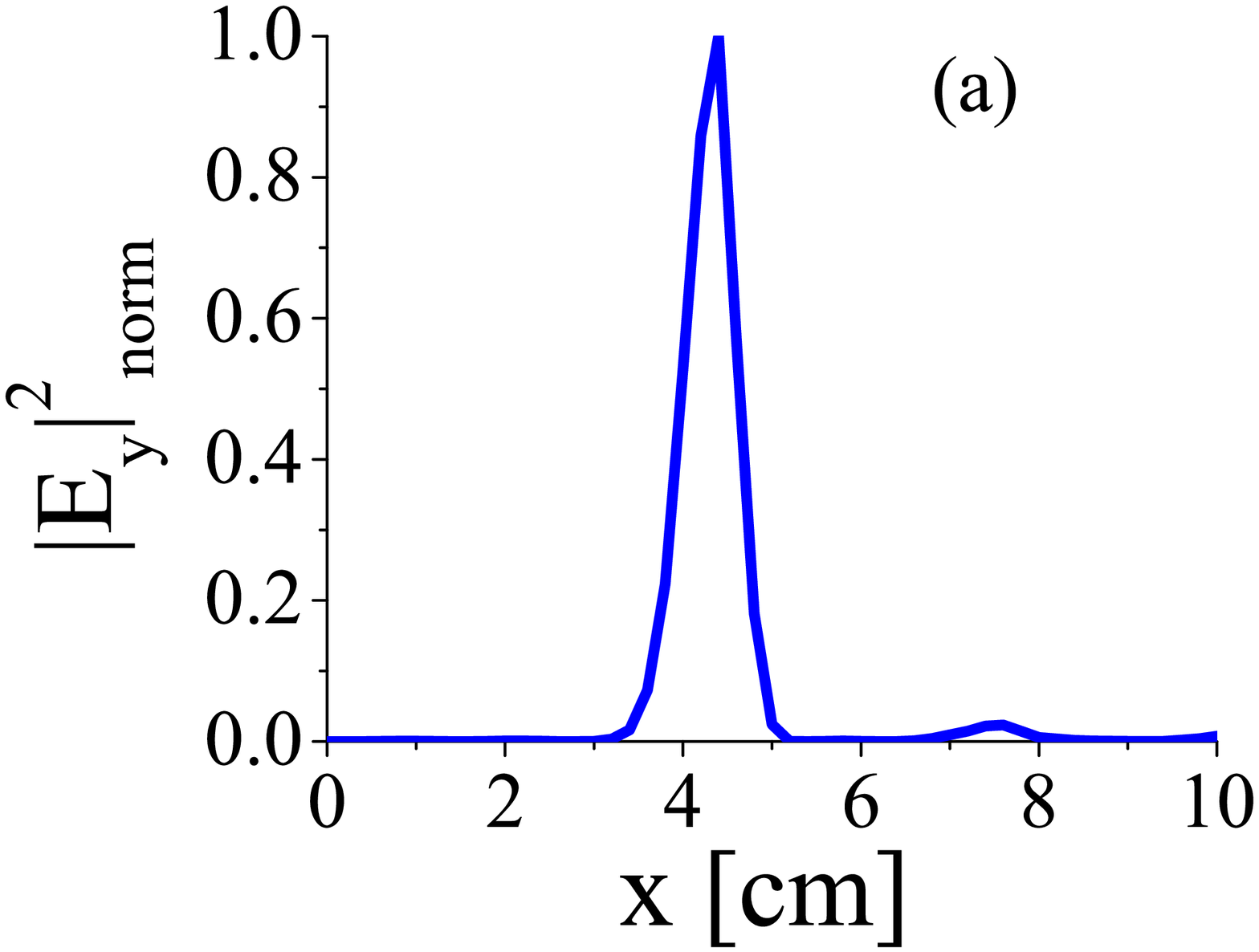}
\hspace{1.2cm}

}
\subfigure{
\includegraphics[width=4.5cm]{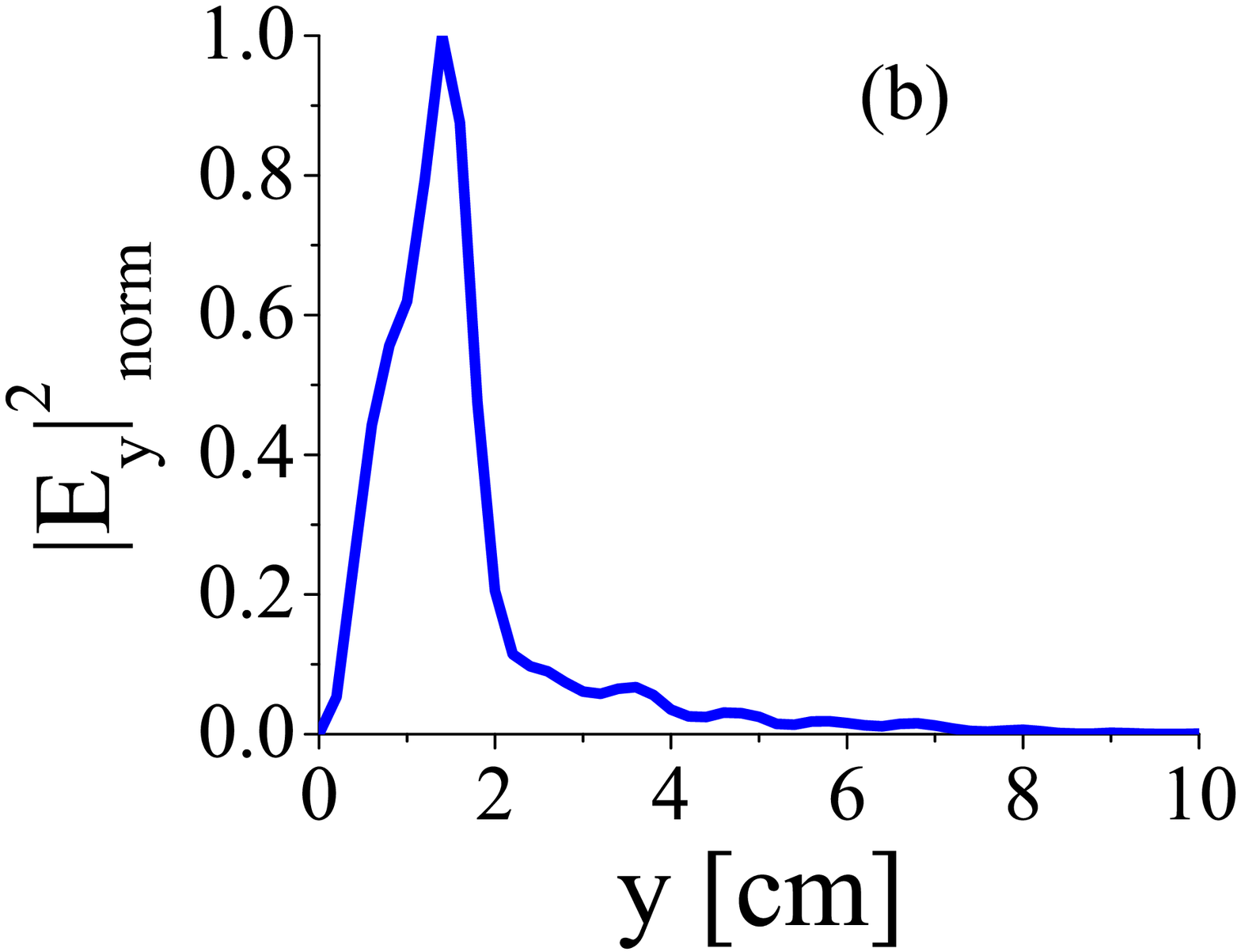}
\hspace{1cm}

}
\caption{ Normalized transversal (a) and lateral (b) experimental profiles, taken at $13.7175 GHz$ along the lines where the focused image exhibits its maximum, for the alumina-air superlens.}

\label{MEAS4}
\end{figure}

In these three bands, the best achieved resolutions (FWHM) are 0.29$\lambda$  at $7.8000 GHz$ ($\omega$ = 0.260), 0.32$\lambda$ at $9.0825 GHz$ ($\omega$ = 0.300)  and 0.29$\lambda$  at $13.7175 GHz$ ($\omega$ = 0.457). Since in Figs. \ref{MEAS2}(a), (c), (e) we use a normalized scale, it is worthwhile to mention that the maximum transmitted intensity decreases with increasing frequency. This is because for higher frequencies the amount of wavevectors collected and focused is lower due to the different shape of the EFCs.

\begin{figure}[htbp!]
\centering
\subfigure{

\includegraphics[width=4.5cm]{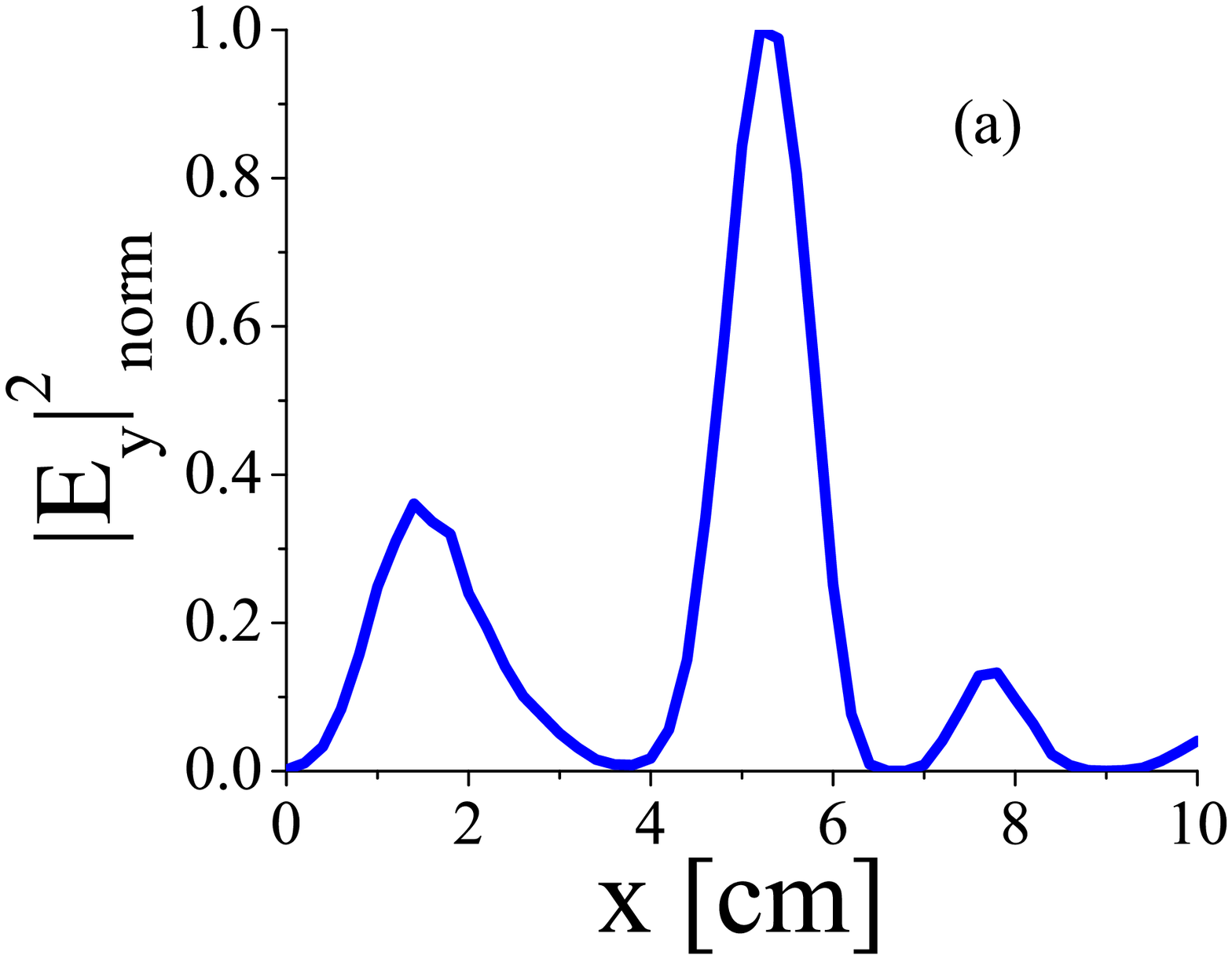}
\hspace{1.2cm}
}
\subfigure{

\includegraphics[width=4.5cm]{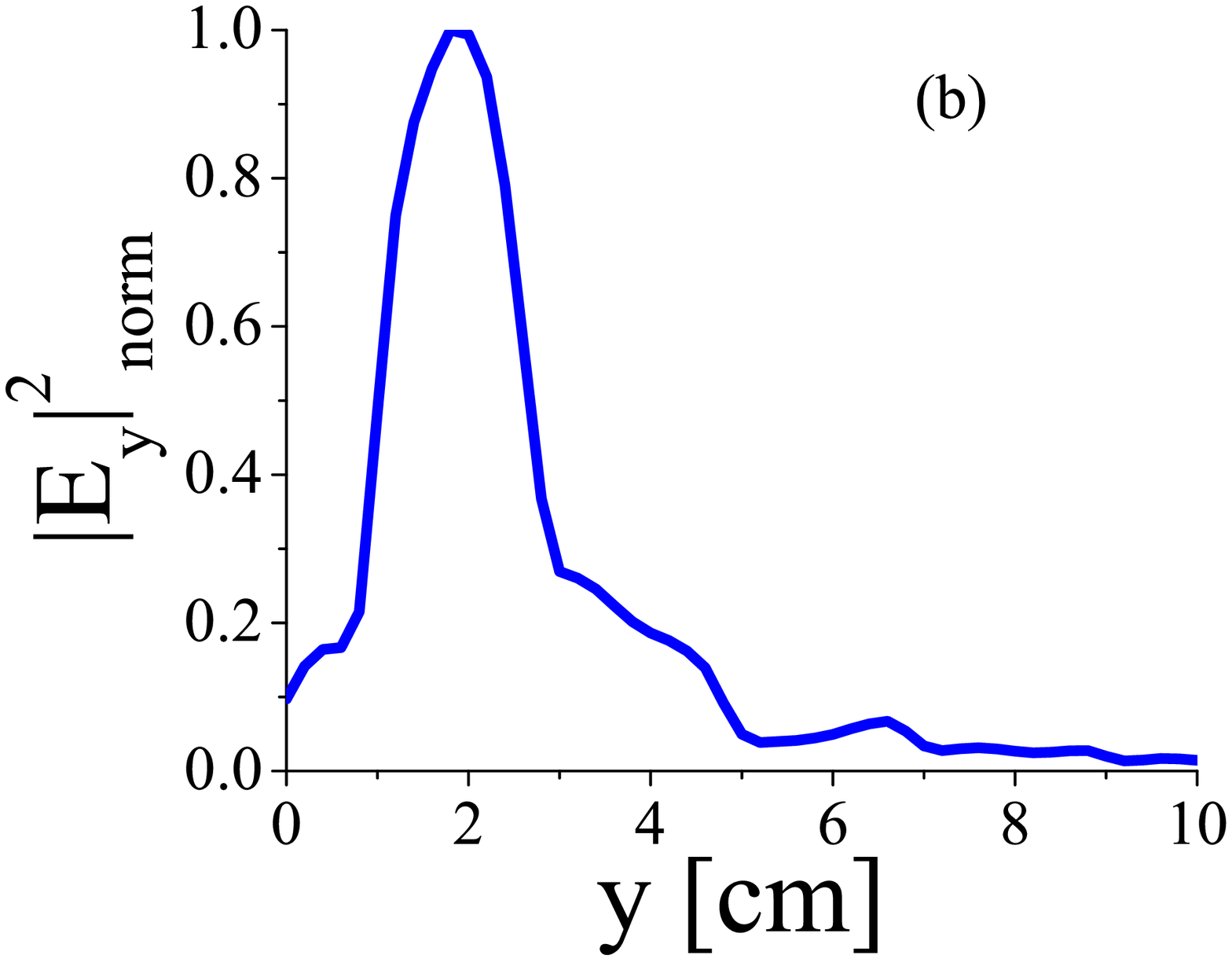}
\hspace{1cm}
}
\caption{ Normalized transversal (a) and lateral (b) experimental profiles, taken at $7.3505 GHz$ along the lines where the focused image exhibits its maximum, for the alumina-plexiglas superlens.}
\label{MEAS1}
\end{figure}

It is also worth noting that the measured values of the FWHM well agree with numerical simulations within 30\%.
In all three cases the focused images are clearly visible, as shown by the spatial mapping of the electric field intensity detected in the image plane and reported in Figs. \ref{MEAS2}(a), (c), and (e). The absence of noticeable aberration is confirmed by the respective phase maps in Figs. \ref{MEAS2}(b), (d), (f), that exhibit a pattern typical of circular waves. In Figs. \ref{MEAS4}(a) and (b) the transversal and lateral profiles respectively for the alumina-air case are shown, as measured at $f = 13.7175 GHz$. In Figs. \ref{MEAS1}(a) and (b)  the transversal and lateral profiles respectively for the alumina-plexiglass case are shown, measured at $f = 7.3505 GHz$ ($\omega$ = 0.245). At this frequency, we obtain the best lens resolution (FWHM) of 0.27$\lambda$.

\begin{figure}[htbp!]
\centering
\subfigure{
\includegraphics[width=6cm]{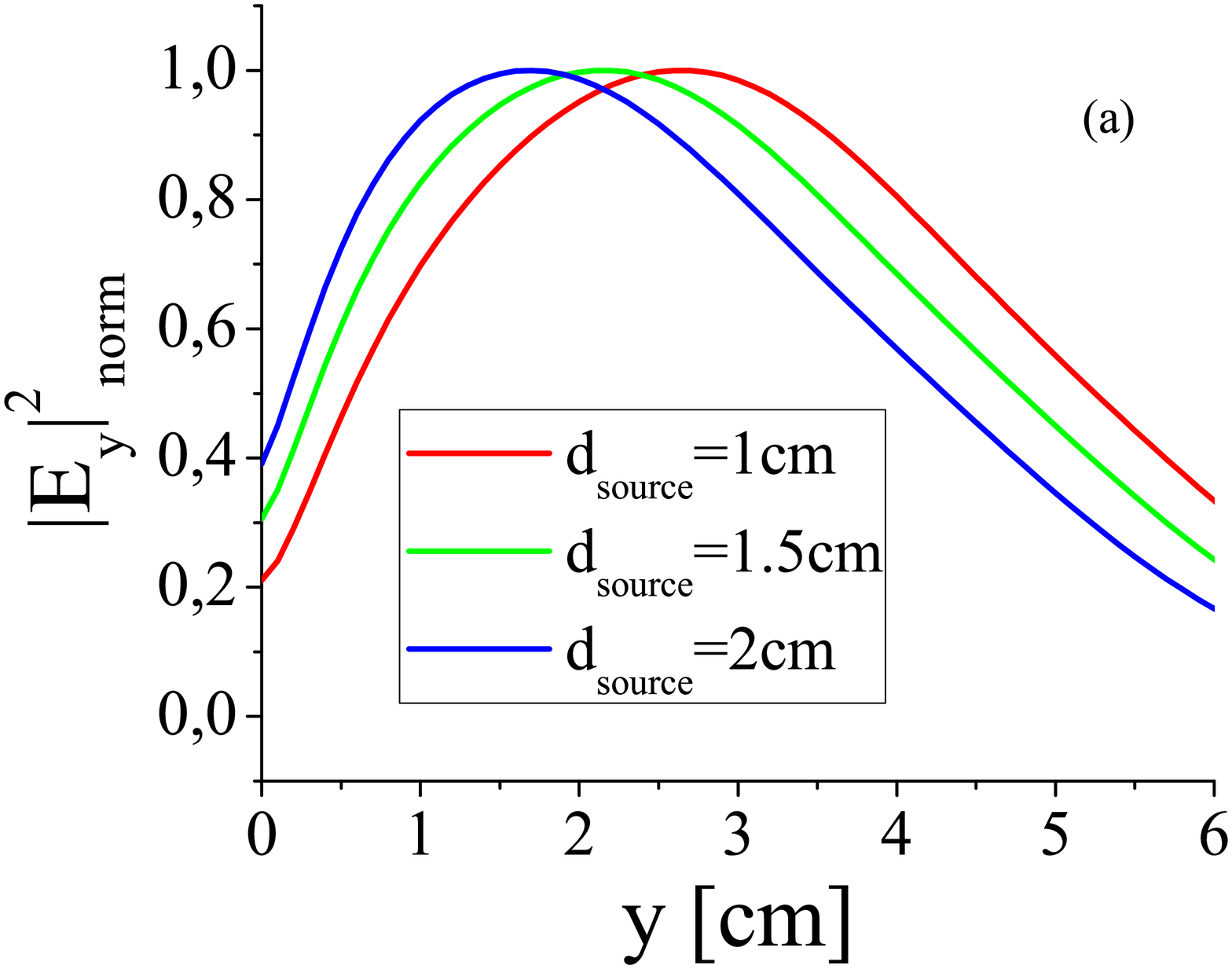}
}
\subfigure{
\includegraphics[width=6.5cm]{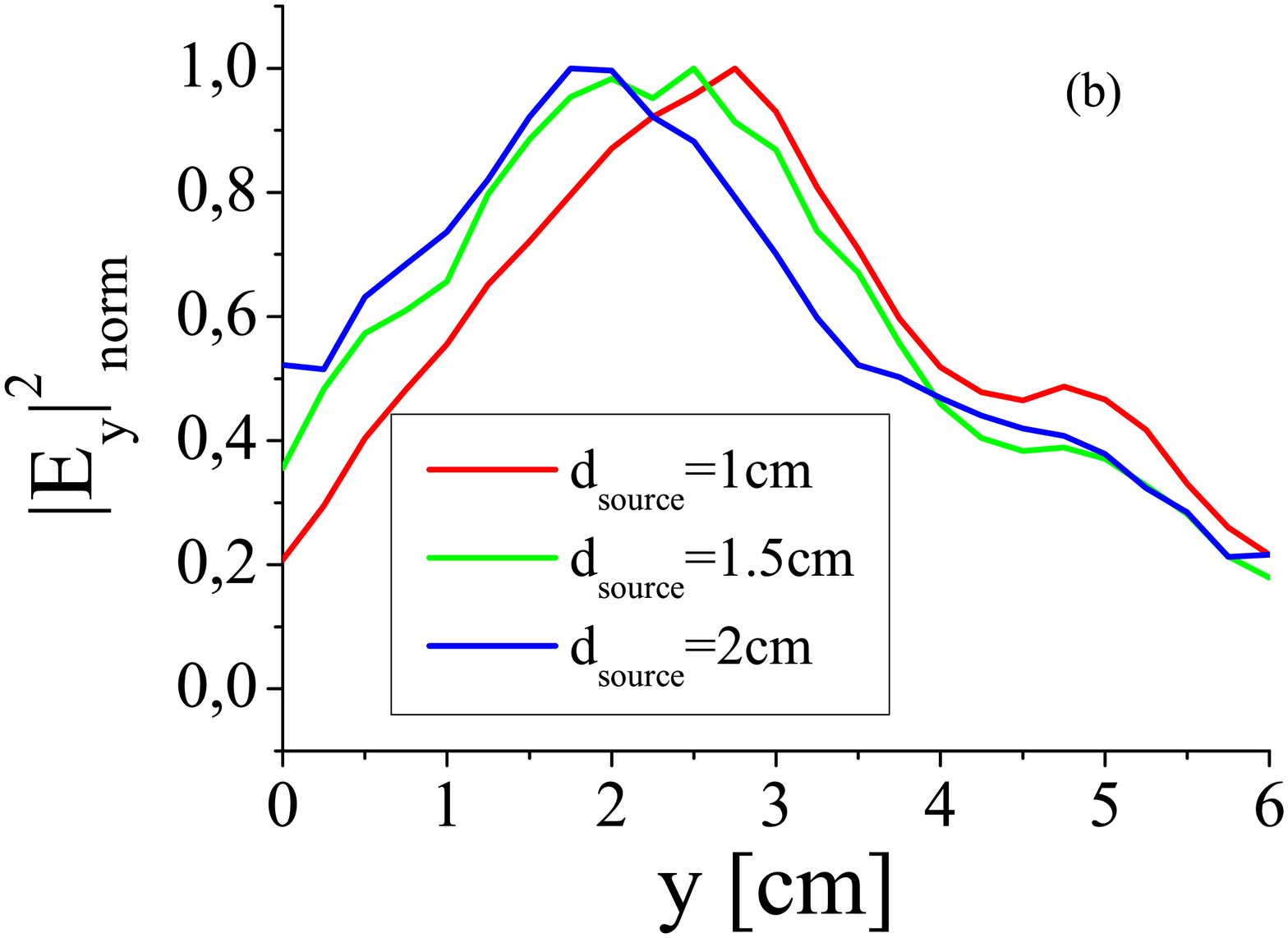}
}
\caption{Simulated (a) and measured (b) lateral profiles respectively for the focused image of a point source changing its position ($d_{source}$) normal to the surface of the alumina-plexiglass PC superlens}
\label{DVAR}
\end{figure}

We also measured the spatial shift of the focused image changing the source position along the \emph{y} direction (normal to the PC surface). In Figures \ref{DVAR}(a) and (b) the simulated and measured lateral profiles respectively of the electric field intensity (in the normalized scale) for the alumina-plexiglass slab are reported. Experimental data (Fig. \ref{DVAR}(b)) clearly show that the image focus moves closer to the PC superlens as long as the point source moves away, strictly following the behavior predicted by the numerical analysis (Fig. \ref{DVAR}(a)), and in accordance also with what expected from a simple ray-diagram.

\section{Conclusions}

We have shown that simple one-dimensional photonic crystals made of tilted dielectric elements with moderately low index contrast may exhibit both positive and negative refraction in multiple frequency bands. This property has been exploited to design and experimentally realize flat slabs capable to produce off-axis subwavelength imaging (superlensing). The lens properties (electric field intensity, transversal and lateral resolution) can be greatly improved by introducing a simple corrugation of the slab surface. In particular, the increase in the amplitude of the signal transmitted through the slab is an important feature to obtain high-resolution imaging of a subwavelength object.

Furthermore, the easiness of lens fabrication, the possibility of using materials with a low index contrast as dielectric elements and the small superlens size render these 1D PC structures an appealing alternative, for specific applications, to 2D or 3D PC slabs.

The natural evolution of the work presented here might be the development of a nano 1D superlens for the near-infrared and optical regimes.

\section{Acknowledgements}

This work was funded by the Italian Ministry of Education and Scientific Research (MIUR) under the PRIN-2006 grant on "Study and realization of metamaterials for electronics and TLC applications". We also thank Sa.Ro s.r.l. and Guazzoni s.r.l. for the supply and the cut of the alumina, and Mr. F. M. Taurino for the technical support.


\begin{thebibliography}{99}


\bibitem{pen} J. B. Pendry, ``Negative Refraction Makes a Perfect Lens,'' \prl {\bf 85,} 3966 (2000).

\bibitem{noto} M. Notomi, ``Theory of light propagation in strongly modulated photonic crystals: refractionlike behavior in the vicinity of the photonic band gap ,'' \prb {\bf 62,} 10696 (2000).

\bibitem{luo1} C. Luo, S. G. Johnson, and J. D. Joannopoulos, ``All-angle negative refraction without negative effective index,'' \prb {\bf 65,} 201104 (2002).

\bibitem{luo2} C. Luo, S. G. Johnson, and J. D. Joannopoulos, ``Subwavelength imaging in photonic crystals,'' \prb,{\bf 68,} 045115 (2003).

\bibitem{par} P. Parimi, W. T. Lu, P. Vodo, and S. Sridhar, ``Imaging by Flat Lens using Negative Refraction,'' \nat {\bf 426,} 404 (2003).

\bibitem{fan1} X. Fan and G. P. Wang, ``Nanoscale metal waveguide arrays as plasmon lenses ,'' \ol {\bf 31 ,} 1322 (2006).

\bibitem{fan2} X. Fan, G. P. Wang, J. C. W. Lee, and C. T. Chan, ``All-Angle Broadband Negative Refraction of Metal Waveguide Arrays in the Visible Range: Theoretical Analysis and Numerical Demonstration ,''  \prl {\bf 97 ,} 073901 (2006).

\bibitem{shi}H. Shin and S. Fan, ``All-angle negative refraction and evanescent wave amplification using one-dimensional metallodielectric photonic crystals ,'' \apl {\bf 89 ,} 151102 (2006).

\bibitem{he} B. Wang, L. Shen, and S. He, ``Superlens formed by a one-dimensional dielectric photonic crystal ,'' \josab {\bf 25,} 391 (2008).

\bibitem{savo}S. Savo, E. Di Gennaro, C. Miletto, A. Andreone, P. Dardano, L.Moretti, and V. Mocella, ``Pendell\"{o}sung effect in photonic crystals,''\opex {\bf 16,} 9097 (2008), \url{http://www.opticsinfobase.org/oe/abstract.cfm?URI=oe-16-12-9097}.

\bibitem{feng} S. Feng, H. Y. Sang, Z. Y. Li, B. Y. Cheng, and D. Z. Zhang, ``Sensitivity of surface states to the stack sequence of one-dimensional photonic crystals ,'' J. Opt. A: Pure Appl. Opt. {\bf 7 ,} 374 (2005).


\bibitem{dec} T. Decoopman, G. Tayeb, S. Enoch, D. Maystre, and B. Gralak, ``Photonic crystal lens: from negative refraction and negative index to negative permittivity and permeability,'' \prl {\bf 97,} 073905 (2006).


\bibitem{mea} R. D. Meade, K. D. Brommer, A. M. Rappe, and J. D. Joannopoulos, ``Electromagnetic Bloch waves at the surface of a photonic crystal,'' \prb {\bf 44,} 10961 (1991).

\bibitem{wang}B. Wang, W. Dai, A. Fang, L. Zhang, G. Tuttle, Th. Koschny, and C. M. Soukoulis, `` Surface waves in photonic crystal slabs ,'' \prb  {\bf 74,} 195104 (2006).

\bibitem{xiao} S. Xiao, M. Qiu, Z. Ruan, and S. He, ``Influence of the surface termination to the point imaging by a photonic crystal slab with negative refraction,'' \apl {\bf 85 ,} 4269 (2004).

\bibitem{mor}E. Moreno, L. Martin-Moreno, and F. J. Garci´a-Vidal, ``Efficient coupling of light into and out of a photonic crystal waveguide via surface modes,'' Photonics and Nanostructures - Fundamentals and Applications {\bf 2,} 97 (2004).

\bibitem{casse}B. D. F. Casse, W. T. Lu, R. K. Banyal, Y. J. Huang, S. Selvarasah, M. R. Dokmeci, C. V.
Perry, and S. Sridhar, ``Imaging with subwavelength resolution by a generalized superlens at infrared wavelengths ,'' \ol, in press).




\end{thebibliography}
\end{document}